\documentclass{IEEEtran}
\usepackage{amsmath}
\usepackage{amssymb}
\usepackage{mathtools}
\usepackage{graphicx}
\usepackage{url}

\usepackage{breakurl}

\DeclareMathOperator*{\per}{Per}

\makeatletter

\begin{document}

\title{Roos' Matrix Permanent Approximation Bounds for Data Association 
Probabilities}

\author{Lingji~Chen%
\thanks{The author is with nuTonomy, Boston, Massachusetts, USA. The research was conducted in part while the author was with BAE Systems. }
}

\maketitle

\begin{abstract}
Matrix permanent plays a key role in data association probability calculations. 
Exact algorithms (such as Ryser's)    scale exponentially with matrix size. 
Fully polynomial time randomized approximation schemes exist but are quite 
complex. This letter introduces to the tracking community a simple 
approximation algorithm with error bounds,  recently developed by Bero Roos, 
and illustrates its potential use for estimating probabilities of data 
association hypotheses.
\end{abstract}

\section{Introduction}
In multi-target tracking, the (normalized) likelihoods of the associations 
between tracks and measurements are calculated using motion and sensor models, 
and for some tracking algorithms, these suffice to define a maximum likelihood 
solution. However, there are situations in which the {\em probabilities} of 
association hypotheses are also important, or even required for the algorithms. 
One example is Joint Probabilistic Data Association Filter (JPDAF) as 
described by \cite{CrouseD17}, where the evaluation of target-measurement 
association probabilities is a necessary part of the algorithm. Another example 
is Generalized Labeled Multi-Bernoulli (GLMB) Filter as described in 
\cite{VoB17}; here the probabilities are not required, but it would be good to 
know, {\em quantitatively}, how much truncation error \cite{VoBN14} has 
occurred: When we keep for example only the top 100 hypotheses, are we keeping 
90\% of the probability mass, or just 50\%?

To get such probabilities we need to normalize by the sum of likelihoods of all 
permissible association  hypotheses, whose number grows combinatorially. If we 
construct a ``likelihood matrix'' whose entries are derived from pairwise 
target-measurement likelihoods, and append it with diagonal matrices for missed 
detections and target deaths, as is done in \cite{VoB17}, then under an 
independence assumption, each hypothesis likelihood is a product of 
``non-conflicting'' terms from this matrix, and the normalizing factor we seek 
is the {\em permanent} of the matrix \cite{wiki:permanent}. 

Exact matrix permanent algorithms, such as Ryser's \cite{RyserH63,CrouseD17}, 
scale exponentially with the matrix size \cite{ValiantL79,LyonsA11}. For a 
matrix with nonnegative entries, a fully polynomial time randomized 
approximation scheme (FPRAS) is presented in \cite{JerrumM04} through Markov 
Chain Monte Carlo (MCMC), which can calculate a solution within a factor of $1 
\pm \epsilon$ of being optimal for a given $\epsilon > 0$. This algorithm is 
quite complex to analyze and implement. On the other hand, as is shown in 
\cite{UhlmannJ04}, even ``crude'' approximations may turn out to be useful for 
estimating various probabilities. With such a motivation, this letter brings to 
the attention of the tracking community a recent result by Bero Roos 
\cite{RoosB17} that provides  a first order and a second order approximations 
to the permanent of a rectangular matrix, both with error bounds\footnote{For higher order approximations, see for example \cite{RoosB15}.}.  

Section~\ref{sec:roos} presents Roos' algorithm with simplified notations. 
Section~\ref{sec:ex} illustrates a potential use for estimating association 
hypotheses probabilities, and also points out the issue with computation time. 
Section~\ref{sec:con} discusses future research.

\section{The Roos' approximations} \label{sec:roos}
We use, for concreteness, the matrix layout in Figure 1 of \cite{VoB17} for 
(normalized) likelihoods\footnote{without taking the logarithm}: Each row 
corresponds to either an existing target, or a potential new-born target from a 
Labeled Multi-Bernoulli birth model. Each column corresponds to one of the 
following situations: (1) a measurement for a survived and detected target, (2) 
a survived but undetected target, and (3) a dead or unborn target. An 
association hypothesis essentially ``picks'' likelihood entries from the matrix, such that 
there is exactly one entry picked for each row, and zero or one entry picked for each 
column. Measurements that are not picked automatically become clutter and 
need not be explicitly dealt with\footnote{which may explain why the 
contemporary  GLMB filters are more efficient than the classical 
Hypothesis-Oriented Multiple Hypothesis Tracker (HO-MHT) \cite{ReidD79}. }. 

Thus the matrix always has more columns than rows. However, in order to follow 
the presentation in \cite{RoosB17} closely, we will describe the algorithm 
using 
a ``thin'' matrix with more rows than columns; this means that we will apply 
Roos' algorithm to the transpose of our likelihood matrix for computation. 

Let $\mathcal{P}_k^n$ denote the set of all $k$-permutations of $n$, the 
ordered arrangements of a $k$-element subset of an $n$-set, and 
$\mathcal{C}_k^n$ denote the set\footnote{To iterate such sets in a memory 
efficient way, see \cite{NijenhuisA78}.} of all $k$-combinations of $n$, the 
unordered $k$-element subset of an $n$-set.  Then the permanent of a thin 
matrix $Z = \left[z_{j,r}\right]_{N \times n}$ is defined as
\begin{equation} \label{eqn:per}
 \per(Z) = \sum_{\sigma \in \mathcal{P}_n^N} \prod_{r=1}^n z_{\sigma(r),r}.
\end{equation}
For $r = 1, \ldots, n$, set the column average to 
\[
 \tilde{z}_r = \frac{1}{N} \sum_{j=1}^N z_{j,r}.
\]
Define for an index subset $S$ the product
\[
 \tilde{p}_S = \prod_{r \in S} \tilde{z}_r.
 \]
Using a Matlab-type notation ``:'' to denote consecutive integers, we define
\[
\tilde{p}^{(1)} = \tilde{p}_{1:n}
\]
and
\[
\tilde{p}^{(2)} = \sum_{R \in \mathcal{C}_2^n} \tilde{p}_{1:n \backslash R} 
\sum_{j=1}^N \prod_{r \in R} (z_{j,r} - \tilde{z}_r),
\]
and state the first and second order approximations respectively as
\[
 \left \vert \per(Z) - \frac{N!}{(N-n)!} \tilde{p}^{(1)} \right \vert \le 
\frac{N!}{(N-n)!} \frac{\theta_2}{2 N} f_2(\sqrt{\beta}, \sqrt{\kappa_2}),
\]
and
\[
\begin{multlined}
 \left \vert  \per(Z) - \left( \frac{N!}{(N-n)!} \tilde{p}^{(1)} - 
\frac{(N-2)!}{(N-n)!} \tilde{p}^{(2)} \right) \right \vert \le \\
 \frac{N!}{(N-n)!} \left( \frac{\theta_3}{2 N^2} f_3(\sqrt{\beta}, 
\sqrt{\kappa_3}) + \frac{\theta_4}{8 N^2} f_4(\sqrt{\beta}, 
\sqrt{\kappa_4})\right),
\end{multlined}
\]
where definitions used in the error bounds are given below. To save space, we 
skip special cases and only describe those where $n\ge 5$.

First, 
\[
\beta = \frac{1}{n} \sum_{r=1}^n |\tilde{z}_r|^2. 
\]
For $\nu = 2, 3, 4$, define
\[
 \kappa_{\nu} = \frac{1}{(n-\nu)(N-\nu)} \max_{J \in \mathcal{C}_\nu^N, R \in 
\mathcal{C}_\nu^n} \sum_{j\in 1:N \backslash J} \; \sum_{r \in 1:n \backslash 
R} \vert z_{j,r} \vert^2.
\]

Define a shorthand notation for row difference
\[
 y_{j,k;r} = z_{j, r} - z_{k, r}, \; j, k \in 1\!:\!N, r \in 1\!:\!n.
\]
Then
\begin{align*}
 \theta_2 &= \alpha_2 \sqrt{\sum_{(r,s) \in \mathcal{P}_2^n } \left( 
\sum_{(u,v) \in \mathcal{P}_2^N} \left \vert y_{u,v;r} y_{u,v;s} \right \vert 
\right)^2 },\\
\theta_3 &= \alpha_3 \sqrt{\sum_{(r,s,t) \in \mathcal{P}_3^n } \left( 
\sum_{(u,v,w) \in \mathcal{P}_3^N} \left \vert y_{u,v;r} y_{u,v;s} y_{u,w;t} 
\right \vert \right)^2 }, \\
\theta_4 &= \alpha_4 \sqrt{\sum_{(q,r,s,t) \in \mathcal{P}_4^n } \left( 
\sum_{(u,v,w,x) \in \mathcal{P}_4^N} \left \vert y_{u,v;q} y_{u,v;r} y_{w,x;s} 
y_{w,x;t} \right \vert \right)^2 },
\end{align*}
where for $d = 2, 3, 4$, the constants are given by 
\[
 \alpha_d = \frac{(N-d)!}{N!} \sqrt{\frac{(n-d)!}{n!}}.
\]
The functions are defined as
\begin{align*}
 f_2(x_1, x_2) &= \sum_{k=2}^n (k-1) x_1^{n-k} x_2^{k-2}, \\
 f_3(x_1, x_2) &= \sum_{k=3}^n (n+k-2) (n-k+1) x_1^{n-k} x_2^{k-3}, \\
 f_4(x_1, x_2) &= \sum_{k=4}^n (k-3)(n+k-2)(n-k+1) x_1^{n-k} x_2^{k-4}.
\end{align*}

\section{An example of ideal usage} \label{sec:ex}
We will illustrate one use of Roos' permanent approximations in the framework 
of GLMB \cite{VoB17}. For ease of exposition we will consider the case where at 
time $k-1$ there is only one hypothesis, and at time $k$ it gives rise to $M$ 
hypotheses, assuming that we enumerate them all. The weight of each hypothesis 
is proportional to the product of likelihoods inside the summation in Equation 
(\ref{eqn:per}), noting that the matrix has been transposed. After normalizing 
the weight by the permanent, we will get the probability of each hypothesis. 

However, for any practical application of GLMB, we cannot enumerate all 
child hypotheses, and have to truncate at a number, say $K$. Then the weights 
are normalized by the sum of these $K$ weights, not the sum of {\em all} 
weights which is given by the permanent. The truncation error is given in 
\cite{VoBN14}, which confirms our intuition that we should pick the highest $K$ 
weights \footnote{or the best $K$ weights we can find within a computation 
budget 
using for example Gibbs sampling} to keep. If we take the negative log of the 
likelihood matrix $Z$, then the best $K$ assignments can be enumerated by the 
Murty's algorithm \cite{MurtyK68,MillerM97}, which calls as a subroutine the 
Munkres algorithm that finds the best bipartite matching 
\cite{MunkresJ57,PilgrimR17,CaoY11}. 

It would be quite useful, even if done offline, to know 
quantitatively what the truncation error is: Do these $K$ hypotheses represent 
90\% of the probability mass, or only 50\%? 

\begin{figure}[!htb]
 \centering
\includegraphics[width=0.99\columnwidth,trim={60 0 40 20}, 
clip]{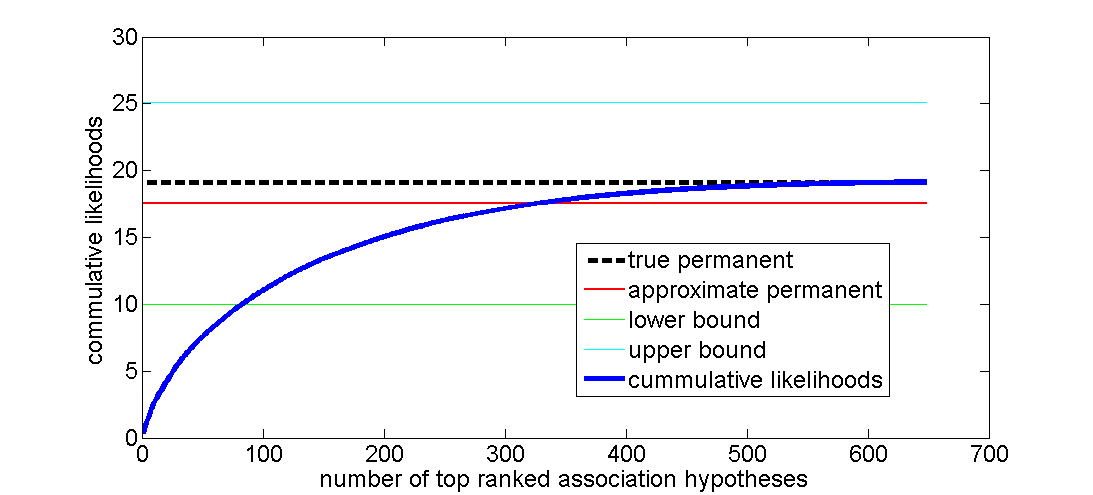}
 \caption{A toy example showing the use of Roos' approximation bounds for 
estimating probabilities of the ``best $K$'' hypotheses.} \label{fig:ex}
\end{figure}

To illustrate the point, we create a toy example with a random likelihood 
matrix 
of size 4 by 12 and run Murty's 
algorithm on it, recording the cummulative likelihoods with each increment of 
the value $K$. This is shown as the blue curve\footnote{If we use Gibbs 
sampling 
instead of Murty's algorithm, the curve will still be increasing but not 
necessarily concave.}
in Figure~\ref{fig:ex}. We also calculate, by Ryser's algorithm, the exact 
permanent and mark it as the black, dotted line. The approximate permanent 
calculated by Roos' second approximation, together with its upper and lower 
bounds, are marked as the red, cyan and green lines.

It can be seen from the figure that if we stop at about the top 25 hypotheses, 
we are guaranteed to capture about $5/25=20\%$ of the total probability, and 
there is even hope that the percentage can be as high as $5/10=50\%$. If we 
continue to obtain the top 100  hypotheses, the lower bound is no longer 
informative, but the upper bound guarantees that we have about  $10/25=40\%$ of 
the probability. Empirically we have observed that Roos' second approximation 
is 
often quite accurate but with conservative bounds, so the percentage $10/18 
\approx 56\%$ may be close to the truth, which we know is $10/19 \approx 53\%$.

\section{The issue of computation time}
\begin{figure}[!htb]
 \centering
\includegraphics[width=0.80\columnwidth,trim={20 0 55 20}, 
clip]{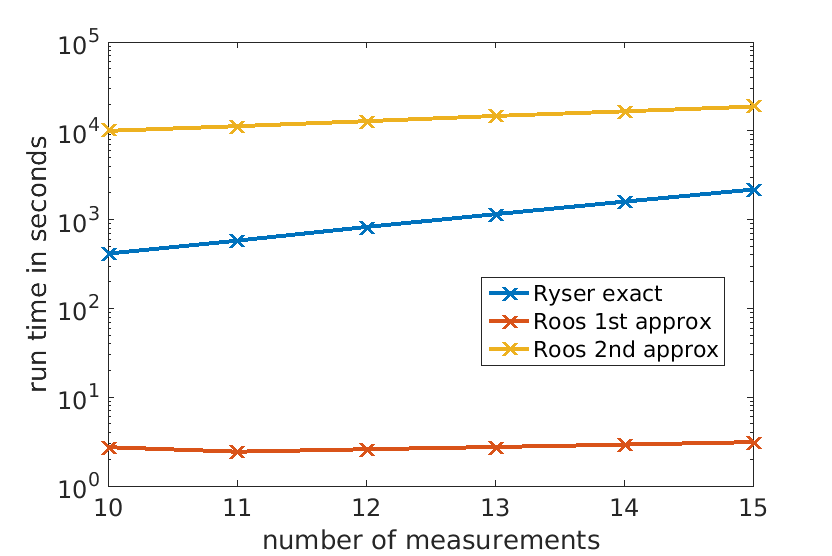}
 \caption{Computation times of unoptimized Matlab code for random likelihood 
matrices with 10 rows and 30 to 35 columns (10 to 15 for the measurement 
block).} \label{fig:times}
\end{figure}

Ryser's algorithm scales exponentially while Roos' approximation scales 
polynomially, so for {\em large} matrices the latter should be faster to 
compute 
than the former. However, for ``mid-sized'' matrices, Roos's first 
approximation 
is fast but conservative while Roos' second approximation is more useful but 
slow, often slower than Ryser's algorithm. This point is illustrated by an 
experiment shown in Figure~\ref{fig:times}, where computation times for random 
likelihood matrices are plotted, based on unoptimized Matlab code. The 
structure 
of the likelihood matrix corresponds to 10 targets (existing and birthing) and 
10 to 15 measurements, i.e., with 10 rows and 30 to 35 columns.

The fact that both Ryser's and Roos' second approximation are unbearably slow 
for a matrix of this size indicate the following possible ways of improvement:
\begin{itemize}
 \item optimized implementation in a compiled languge;
 \item parallelized implementation;
 \item exploitation of the diagonal structure of the second and third blocks of 
the likelihood matrix (used in GLMB filtering);
 \item better approximation algorithms.
\end{itemize}

\section{Conclusions} \label{sec:con}
In this letter we have presented Roos' approximation algorithms with bounds for 
matrix permanent. We illustrated their use in estimating data association 
probablities, such as in GLMB filtering where only the top $K$ hypotheses are 
kept. We pointed out the challenge in computation time and proposed directions 
for improvements. 

\section*{Acknowledgment}
The author would like to thank Professor Bero Roos for discussions and 
clarifications, and Professor Jeffrey Ulhmann for providing a cited paper. He 
would also like to thank Ms. Emily Polson for support.


\def\url#1{}
\raggedright
\bibliographystyle{myIEEEtran}
\bibliography{taes18}

\begin{thebibliography}{10}
\providecommand{\url}[1]{#1}
\csname url@samestyle\endcsname
\providecommand{\newblock}{\relax}
\providecommand{\bibinfo}[2]{#2}
\providecommand{\BIBentrySTDinterwordspacing}{\spaceskip=0pt\relax}
\providecommand{\BIBentryALTinterwordstretchfactor}{4}
\providecommand{\BIBentryALTinterwordspacing}{\spaceskip=\fontdimen2\font plus
\BIBentryALTinterwordstretchfactor\fontdimen3\font minus
  \fontdimen4\font\relax}
\providecommand{\BIBforeignlanguage}[2]{{%
\expandafter\ifx\csname l@#1\endcsname\relax
\typeout{** WARNING: IEEEtran.bst: No hyphenation pattern has been}%
\typeout{** loaded for the language `#1'. Using the pattern for}%
\typeout{** the default language instead.}%
\else
\language=\csname l@#1\endcsname
\fi
#2}}
\providecommand{\BIBdecl}{\relax}
\BIBdecl

\bibitem{CrouseD17}
\BIBentryALTinterwordspacing
D.~F. Crouse and P.~Willett, ``{Computation of Target-Measurement Association
  Probabilities Using the Matrix Permanent},'' \emph{IEEE Transactions on
  Aerospace and Electronic Systems}, vol.~53, no.~2, pp. 698--702, Apr. 2017.
  \url{http://dx.doi.org/10.1109/taes.2017.2664479}
\BIBentrySTDinterwordspacing

\bibitem{VoB17}
\BIBentryALTinterwordspacing
B.~N. Vo, B.-T. Vo, and H.~Hoang, ``{An Efficient Implementation of the
  Generalized Labeled Multi-Bernoulli Filter},'' \emph{IEEE Transactions on
  Signal Processing}, vol.~65, no.~8, pp. 1975--1987, Apr. 2017.
  \url{http://dx.doi.org/10.1109/tsp.2016.2641392}
\BIBentrySTDinterwordspacing

\bibitem{VoBN14}
\BIBentryALTinterwordspacing
B.-N. Vo, B.-T. Vo, and D.~Phung, ``{Labeled Random Finite Sets and the Bayes
  Multi-Target Tracking Filter},'' \emph{Signal Processing, IEEE Transactions
  on}, vol.~62, no.~24, pp. 6554--6567, Dec. 2014.
  \url{http://dx.doi.org/10.1109/tsp.2014.2364014}
\BIBentrySTDinterwordspacing

\bibitem{wiki:permanent}
\BIBentryALTinterwordspacing
Wikipedia, ``Permanent -- {W}ikipedia{,} the free encyclopedia,'' 2018,
  [Online; accessed 16-February-2018 ].
  \url{https://en.wikipedia.org/w/index.php?title=Permanent\&\#38;oldid=795641630}
\BIBentrySTDinterwordspacing

\bibitem{RyserH63}
\BIBentryALTinterwordspacing
H.~J. Ryser, \emph{{Combinatorial mathematics}}, ser. Carus mathematical
  monographs.\hskip 1em plus 0.5em minus 0.4em\relax Mathematical Association
  of America; distributed by Wiley [New York], 1963.
  \url{https://books.google.com/books?id=wOruAAAAMAAJ}
\BIBentrySTDinterwordspacing

\bibitem{ValiantL79}
\BIBentryALTinterwordspacing
L.~G. Valiant, ``{The complexity of computing the permanent},''
  \emph{Theoretical Computer Science}, vol.~8, no.~2, pp. 189--201, Jan. 1979.
  \url{http://dx.doi.org/10.1016/0304-3975(79)90044-6}
\BIBentrySTDinterwordspacing

\bibitem{LyonsA11}
A.~Lyons, ``{Polynomial-Time Approximation of the Permanent},'' \emph{Course
  project for MATH}, vol. 100, 2011.

\bibitem{JerrumM04}
\BIBentryALTinterwordspacing
M.~Jerrum, A.~Sinclair, and E.~Vigoda, ``{A Polynomial-time Approximation
  Algorithm for the Permanent of a Matrix with Nonnegative Entries},'' \emph{J.
  ACM}, vol.~51, no.~4, pp. 671--697, Jul. 2004.
  \url{http://dx.doi.org/10.1145/1008731.1008738}
\BIBentrySTDinterwordspacing

\bibitem{UhlmannJ04}
\BIBentryALTinterwordspacing
J.~K. Uhlmann, ``{Matrix permanent inequalities for approximating joint
  assignment matrices in tracking systems},'' \emph{Journal of the Franklin
  Institute}, vol. 341, no.~7, pp. 569--593, Nov. 2004.
  \url{http://dx.doi.org/10.1016/j.jfranklin.2004.07.003}
\BIBentrySTDinterwordspacing

\bibitem{RoosB17}
\BIBentryALTinterwordspacing
B.~Roos, ``{New permanent approximation inequalities via identities},''
  \emph{arXiv preprint arXiv:1612.03702}, 2017.
  \url{https://arxiv.org/pdf/1612.03702.pdf}
\BIBentrySTDinterwordspacing

\bibitem{RoosB15}
\BIBentryALTinterwordspacing
------, ``{On Bobkovas approximate de Finetti representation via approximation
  of permanents of complex rectangular matrices},'' \emph{Proceedings of the
  American Mathematical Society}, vol. 143, no.~4, pp. 1785--1796, 2015.
  \url{http://dx.doi.org/10.1090/s0002-9939-2014-12429-4}
\BIBentrySTDinterwordspacing

\bibitem{ReidD79}
\BIBentryALTinterwordspacing
D.~Reid, ``{An algorithm for tracking multiple targets},'' \emph{IEEE
  Transactions on Automatic Control}, vol.~24, no.~6, pp. 843--854, Dec. 1979.
  \url{http://dx.doi.org/10.1109/tac.1979.1102177}
\BIBentrySTDinterwordspacing

\bibitem{NijenhuisA78}
\BIBentryALTinterwordspacing
A.~Nijenhuis, W.~Rheinboldt, and H.~S. Wilf, \emph{{Combinatorial
  Algorithms}}.\hskip 1em plus 0.5em minus 0.4em\relax Academic Press, 1978.
  \url{http://www.worldcat.org/isbn/9780125192606}
\BIBentrySTDinterwordspacing

\bibitem{MurtyK68}
\BIBentryALTinterwordspacing
K.~G. Murty, ``{Letter to the Editor—An Algorithm for Ranking all the
  Assignments in Order of Increasing Cost},'' \emph{Operations Research},
  vol.~16, no.~3, pp. 682--687, 1968.
  \url{http://dx.doi.org/10.1287/opre.16.3.682}
\BIBentrySTDinterwordspacing

\bibitem{MillerM97}
\BIBentryALTinterwordspacing
M.~L. Miller, H.~S. Stone, and I.~J. Cox, ``{Optimizing Murty's ranked
  assignment method},'' \emph{IEEE Transactions on Aerospace and Electronic
  Systems}, vol.~33, no.~3, pp. 851--862, Jul. 1997.
  \url{http://dx.doi.org/10.1109/7.599256}
\BIBentrySTDinterwordspacing

\bibitem{MunkresJ57}
\BIBentryALTinterwordspacing
J.~Munkres, ``{Algorithms for the Assignment and Transportation Problems},''
  \emph{Journal of the Society for Industrial and Applied Mathematics}, vol.~5,
  no.~1, pp. 32--38, Mar. 1957.  \url{http://dx.doi.org/10.1137/0105003}
\BIBentrySTDinterwordspacing

\bibitem{PilgrimR17}
R.~A. Pilgrim, ``{Munkres' Assignment Algorithm Modified for Rectangular
  Matrices},'' http://csclab.murraystate.edu/\~{}bob.pilgrim/445/munkres.html,
  2017.

\bibitem{CaoY11}
Y.~Cao, ``{munkres.m},'' 2011, [Downloaded from Matlab Central,
  http://goo.gl/9YPMi7].

\end{thebibliography}


\end{document}